\documentclass[%
 reprint,
 superscriptaddress,
 amsmath,amssymb,
 aps,
 multiline,
 prl,
]{revtex4-2}
\usepackage{graphicx} 
\usepackage{xcolor}
\usepackage{amsmath}
\usepackage{amssymb}
\usepackage[ruled,vlined]{algorithm2e}
\usepackage{multirow}
\usepackage{rotating}

\renewcommand{\selectlanguage}[1]{}

\begin{document}

\title{Explainable physics-based constraints on reinforcement learning for accelerator controls}

\date{\today}

\author{Jonathan Colen}
\affiliation{Joint Institute on Advanced Computing for Environmental Studies, Old Dominion University, Norfolk, Virginia 23539 USA}
\affiliation{Hampton Roads Biomedical Research Consortium, Portsmouth, Virginia 23703 USA}
\author{Malachi Schram }
\affiliation{Joint Institute on Advanced Computing for Environmental Studies, Old Dominion University, Norfolk, Virginia 23539 USA}
\affiliation{Data Science Department, Thomas Jefferson National Accelerator Facility, Newport News, VA 23606 USA}
\affiliation{Computer Science Department, Old Dominion University, Norfolk, Virginia 23539 USA}
\author{Kishansingh Rajput }
\affiliation{Data Science Department, Thomas Jefferson National Accelerator Facility, Newport News, VA 23606 USA}
\affiliation{Department of Computer Science, University of Houston, Houston, TX 77204, USA}
\author{Armen Kasparian}
\affiliation{Data Science Department, Thomas Jefferson National Accelerator Facility, Newport News, VA 23606 USA}

\begin{abstract}
    We present a reinforcement learning (RL) framework for controlling particle accelerator experiments that builds explainable physics-based constraints on agent behavior. 
    The goal is to increase transparency and trust by letting users verify that the agent's decision-making process incorporates suitable physics. 
    Our algorithm uses a learnable surrogate function for physical observables, such as energy, and uses them to fine-tune how actions are chosen. 
    This surrogate can be represented by a neural network or by an interpretable sparse dictionary model.
    We test our algorithm on a range of particle accelerator controls environments designed to emulate the Continuous Electron Beam Accelerator Facility (CEBAF) at Jefferson Lab.
    By examining the mathematical form of the learned constraint function, we are able to confirm the agent has learned to use the established physics of each environment. 
    In addition, we find that the introduction of a physics-based surrogate enables our reinforcement learning algorithms to reliably converge for difficult high-dimensional accelerator controls environments. 
\end{abstract}

\maketitle

\section{Introduction}

Large-scale scientific systems are difficult to operate, requiring simultaneous control of many apparatus to achieve experimental objectives. To be successful, an operator must understand both the scientific principles of the device as well as the subtle ways that those principles and machine idiosyncrasies collectively influence operation. Data-driven techniques such as reinforcement learning~\cite{sutton_reinforcement_2020} are promising tools to address these challenges, with demonstrated success at playing games~\cite{silver_general_2018,kaufmann_champion-level_2023}, solving engineering problems~\cite{manuel_davila_delgado_robotics_2022,kober_reinforcement_2013} and even scientific tasks~\cite{kain_sample-efficient_2020,kaiser_reinforcement_2024,falk_learning_2021,banerjee_survey_2023}. 
However, these tools remain a black box. While deep neural networks can make accurate predictions, less is known about how to interpret their behavior and understand the rules that govern their operation. This limited explainability hinders trust and is a barrier to applying these tools, especially in environments where unexpected behavior and failures are expensive. 

Physics-informed learning methods introduce physical constraints to improve performance and guide predictions to obey established governing principles~\cite{raissi_physics-informed_2019,karniadakis_physics-informed_2021,banerjee_survey_2023}. However, they require that the system adheres closely to the imposed model and can break down for problems whose laws are not firmly established or in experimental systems where hardware, implementations, and noise may obscure the physical rules or even the relevant variables~\cite{schmitt_machine_2024,seara_sociohydrodynamics_2024}. Symbolic and sparse regression aim to infer these laws directly from data~\cite{schmidt_distilling_2009,brunton_discovering_2016,brunton_machine_2020} and have been incorporated in machine learning pipelines for model discovery~\cite{champion_data-driven_2019,champion_unified_2020,schmitt_machine_2024,supekar_learning_2023,joshi_data-driven_2022,golden_physically_2023,colen_interpreting_2024}. Recent work has explored how to integrate such methods into reinforcement learning schemes by building explainable surrogates of the learned policy, environment, and reward functions~\cite{zolman_sindy-rl_2024}. 
However, these surrogates may be difficult to interpret, particularly for problems where the state-action-reward relationship involves additional details beyond the underlying fundamental physics.

In this work, we present a reinforcement learning (RL) framework that uses learnable and explainable constraints to improve control of particle accelerator experiment environments.
We use trainable surrogate models to identify the dynamical rules for relevant physical observables such as energy. 
These models can be deep neural networks or sparse dictionary models, which provide the additional advantage of learning an interpretable mathematical form for these observables. 
The constraint models train alongside a traditional actor-critic-style model and may influence the chosen actions. 
We test our algorithms on surrogate implementations of linear accelerators (linacs) in Jefferson Lab's Continuous Electron Beam Accelerator Facility (CEBAF)~\cite{adderley_continuous_2024}. 
For simpler problems, such as controlling a single cryomodule, our framework achieves comparable performance to a standard model-free RL algorithm.
For the more challenging task of controlling the entire CEBAF north linac, our framework achieves superior performance at minimizing operational hazards while maintaining a target energy gain.
Our learnable and explainable constraint models also provide a window inside the black box, allowing operators to verify that agent decision-making is based on a correct understanding of the underlying physics.
Our algorithm represents a promising step towards the use of RL to control complex experimental apparatus and enable scientific discovery.

\section{Previous Work}

Machine learning (ML) has been applied to a wide range of problems in physics and accelerator applications. Several studies have applied deep learning models to detect anomalies and operational faults at different experimental sites~\cite{marcato_machine_2021,alanazi_multi-module-based_2023,edelen_anomaly_2021,blokland_uncertainty_2022}. Other efforts have aimed to tune beam properties to reach desired experimental conditions~\cite{roussel_multiobjective_2021,scheinker_model-independent_2013,scheinker_online_2020,kaiser_reinforcement_2024,kain_sample-efficient_2020}. A recent analysis~\cite{rajput_harnessing_2024} compared the performance of Bayesian optimization~\cite{roussel_multiobjective_2021}, and genetic algorithms~\cite{terzic_simultaneous_2014} against RL techniques and found that deep differentiable RL~\cite{jaisson_deep_2022} achieved excellent performance on CEBAF beam optimization. A drawback of this approach is that it requires a fully differentiable environment. Moreover, it remains a black box and gives operators little insight into the rules governing predicted operational behavior.

In physics applications, a path to explainability comes from the knowledge that systems are constrained to obey physical laws. Many studies have examined how to discover these laws using ML techniques~\cite{schmidt_distilling_2009}. Sparse model regression solves this task by selecting a simple and parsimonious model from a dictionary of candidate expressions~\cite{brunton_discovering_2016}. This method has been integrated with deep learning architectures to derive predictive coordinates that are well-suited to sparse modeling~\cite{champion_data-driven_2019}. Other techniques use information-theoretic frameworks, such as the information bottleneck~\cite{tishby_information_2000}, to distill reduced order models from data~\cite{murphy_machine-learning_2024,murphy_information_2024,schmitt_information_2024,koch-janusz_mutual_2018,gokmen_statistical_2021}. The development of these physics-inspired ML frameworks has enabled physical model discovery in fields such as materials~\cite{schoenholz_structural_2016,zhang_analyses_2022,supekar_learning_2023,joshi_data-driven_2022,golden_physically_2023,colen_interpreting_2024} and biology~\cite{mangan_inferring_2016,soelistyo_learning_2022,schmitt_machine_2024,romeo_learning_2021}. 
These techniques have begun to be extended to RL. A recent study integrated sparse model regression with model-based RL, learning parsimonious models of the system and agent from episodic observations~\cite{zolman_sindy-rl_2024}. However, the learned surrogates, which were sufficiently accurate for learning purposes, did not yield ready interpretations or comparisons to known physics. 

\section{Methods}

\subsection{CEBAF environment description}

CEBAF is the primary accelerator at Thomas Jefferson National Accelerator Facility (JLab)~\cite{adderley_continuous_2024}. It accelerates electrons using a pair of superconducting radiofrequency (SRF) linear accelerators containing a series of individually-tunable SRF cavities. During an experiment, the operator must set the gradients in each cavity in order to obtain the target energy gain. At the same time, they aim to avoid potential problems, such as a high heat load or the creation of Fast Shut Down (FSD) trips. To accomplish these goals, one has to consider both the relevant physics as well as the unique operating characteristics of each cavity.

To learn to control CEBAF and set SRF cavity gradients, we used a computational surrogate environment for CEBAF which has been described previously~\cite{rajput_harnessing_2024}. 
In this environment, operating agents observe a state vector containing the current gradients set for each SRF cavity, and take actions to change those gradients to new values. 
During operation, the cavities dissipate heat into the system which depends on a number of factors and is captured by the following equation~\cite{benesch_longitudinal_2015}
\begin{align}
    H = \sum_i \frac{G_i^2 \, \ell_i }{\omega_i \, Q_i(G_i) }
    \label{eq:heat_load}
\end{align}
Here $i$ runs over all cavities,  $G_i$ is the cavity gradient, and the parameters $\ell_i, \omega_i, Q_i$ are the length, impedance, and quality factor of each cavity. 

CEBAF contains a number of legacy cryomodules which can trigger arc faults during operation. When an arc fault is detected, the cavity powers down and the electron beam production halts to prevent damage. 
A previous study characterized the rate of these FSD trips for each cavity (trip rate) using the following statistical model~\cite{benesch_longitudinal_2015}
\begin{align}
    T = \sum_i \exp \left[ {A + B_i(G_i - F_i)} \right]
    \label{eq:trip_rate}
\end{align}
Here $A, B_i$ are regression parameters and $F_i$ is a fault gradient fit from historical data.

The configuration of cavity gradients also determines the total energy gain within the linac. During experiments, an operator aims to keep this energy gain within a narrow range centered at a target energy given by experimental requirements. 
The total energy gain is given by the equation
\begin{align}
    E = \sum_i G_i \ell_i
    \label{eq:energy_gain}
\end{align}

We examined four CEBAF optimization problems in this study. Three are test problems for controlling one, two, or four cryomodules. The final problem tasks the agent with controlling the full CEBAF north linac. The problem sizes and energy targets are summarized in Table~\ref{tab:cebaf_problems}.

\begin{table}[t]
    \centering
    \begin{tabular}{c| c | c | c}
         Problem & Num. Cavities  & $E_{\text{target}}$ (MeV) & $\delta E$ (MeV) \\
         \hline
         \hline
         8D & 8 & 20.08 & 0.40 \\
         16D & 16 & 50.00 & 0.60 \\
         32D & 32 & 120.00 & 0.80 \\
         North linac & 197 & 1050.00 & 2.00 
    \end{tabular}
    \caption{CEBAF optimization problems}
    \label{tab:cebaf_problems}
\end{table}

\subsection{Reinforcement learning algorithms}

In this study, we examined the ability of reinforcement learning (RL) algorithms to operate and control the CEBAF accelerator. 
In a standard RL problem, an agent operates within an environment and obtains rewards based on how well it performs some predetermined task. 
At each time step, the agent observes a state $s$ which is a vector of measurable quantities within the environment. Based on that state, the agent selects a suitable action $a$. The environment uses that action to transition to the next state $s'$ and gives the agent a reward $r$. The goal of the agent is to learn to select the action that maximizes the cumulative reward received for both current and future actions.

\begin{algorithm}[t]
\caption{TD3 + learnable constraints (LC-TD3)}
\label{alg:lctd3}
\SetAlgoLined
Initialize critics $Q_{\theta_1}, Q_{\theta_2}$ and policy network $\pi_{\phi}$ \\
Initialize target networks $\theta_{i'} \gets \theta_{i}$, $\phi' \gets \phi$ \\
Initialize replay buffer $\mathcal{B}$ \\
Initialize learnable surrogate network $O_{\xi}$ and constraint function $C(o)$ \\
\For{$e$ in $1\dots N_e$}{
    Observe state $s$ and select action $a \sim \pi_{\phi}$ \\
    Execute $a$ in environment \\
    Observe next state $s'$, reward $r$, terminal signal $d$, and environmental observables $o$ \\
    Store $(s, a, r, s', d, o)$ in replay buffer $\mathcal{B}$ \\
    \If{time to update}{
        Sample batch of transitions $b \sim \mathcal{B}$ \\
        $a' \gets \pi_{\phi'}(s') + \epsilon\, \mathcal{N}(0, \sigma)$ \\
        $y_i \gets r + \gamma \min_i Q_{\theta'_i}(s', a')$ \\
        Update surrogate $O_{\xi}$ with gradient descent using  
        $\frac{1}{|b|} \nabla_{\xi} \sum \left( O_{\xi}(s, a) - o\right)^2$ \\
        Update $Q$ functions with gradient descent using 
        $\frac{1}{|b|} \nabla_{\theta_i} \sum \left( Q_{\theta_i}(s, a) - y_i \right)^2$ \\
        Update policy $\pi$ with gradient ascent using 
        $\frac{1}{|b|} \nabla_{\phi} \sum \left( Q_{\phi_1}(s, \pi_{\phi}(s)) - \beta\, C( O_{\xi}( s, \pi_{\phi}(s)) \right) $ \\
        Update target networks: \\
        $\theta_i' \gets \tau \theta_i' + (1 - \tau) \theta_i $ \\
        $ \phi' \gets \tau \phi' + (1 - \tau) \phi$
    }
}
\end{algorithm}
\begin{algorithm}[t]
\caption{TD3 + sparse learnable constraints (Sparse LC-TD3)}
\label{alg:sindy-lctd3}
\SetAlgoLined
Initialize critics $Q_{\theta_1}, Q_{\theta_2}$ and policy network $\pi_{\phi}$ \\
Initialize target networks $\theta_{i'} \gets \theta_{i}$, $\phi' \gets \phi$ \\
Initialize replay buffer $\mathcal{B}$ \\
Initialize library function $L(s, a)$, weight vector $\mathbf{w}$, and constraint function $C(o)$ \\
\For{$e$ in $1\dots N_e$}{
    Observe state $s$ and select action $a \sim \pi_{\phi}$ \\
    Execute $a$ in environment \\
    Observe next state $s'$, reward $r$, terminal signal $d$, and environmental observables $o$ \\
    Store $(s, a, r, s', d, o)$ in replay buffer $\mathcal{B}$ \\
    \If{time to update}{
        Sample batch of transitions $b \sim \mathcal{B}$ \\
        $a' \gets \pi_{\phi'}(s') + \epsilon\, \mathcal{N}(0, \sigma)$ \\
        $y_i \gets r + \gamma \min_i Q_{\theta'_i}(s', a')$ \\
        Update weight vector $\mathbf{w}$ with gradient descent using 
        $\frac{1}{|b|} \nabla_{\mathbf{w}} \sum \left( L(s, a) \mathbf{w} - o\right)^2$ \\
        Update $Q$ functions with gradient descent using 
        $\frac{1}{|b|} \nabla_{\theta_i} \sum \left( Q_{\theta_i}(s, a) - y_i \right)^2$ \\
        Update policy $\pi$ with gradient ascent using 
        $\frac{1}{|b|} \nabla_{\phi} \sum \left( Q_{\phi_1}(s, \pi_{\phi}(s)) - \beta\, C( L(s, \pi_{\phi}(s)) \mathbf{w}) \right) $ \\
        Update target networks: \\
        $\theta_i' \gets \tau \theta_i' + (1 - \tau) \theta_i $ \\
        $ \phi' \gets \tau \phi' + (1 - \tau) \phi$
    }
}
\end{algorithm}

A wide variety of techniques have been proposed for RL problems~\cite{sutton_reinforcement_2020}. Here, we considered the Twin-Delayed Deep Deterministic Policy Gradient (TD3) algorithm~\cite{fujimoto_addressing_2018}, which uses one policy model to select actions $\pi(s)$ and a pair of Q-functions $Q(s, a)$ that evaluate the expected future reward resulting from a given action. During training, the agent aims to explore the environment such that the Q-functions faithfully represent the reward space and adjusts the policy to select actions that maximize $Q(s, \pi(s))$. The complete algorithm is provided for reference in Alg.~\ref{alg:td3}. 

From the problem description given previously, we modified the RL problem to include constraints $C(o)$ that operate on a set of physical observables $o$ that are returned by the environment alongside the reward $r$ and next state $s'$. 
For CEBAF, these observables are the energy gain and the constraint objective is the energy target $ | E_{\text{target}} - \sum_i G_i \ell_i | < \delta E$.
The RL agent must learn to configure the cavity gradients to minimize operational hazards such as heat load and FSD trip rate while satisfying this energy constraint. 
With the standard TD3 algorithm, one could modify the reward to promote adherence to the constraint function. However, recent work has found that traditional TD3 struggles to adapt to high-dimensional accelerator environments with hard energy constraints~\cite{rajput_harnessing_2024}.

We modified the TD3 algorithm by introducing a learnable surrogate function that maps state-action pairs to the relevant physical observables $o$.
We considered two cases: the first uses a deep neural network to learn the surrogate function, while the second uses a sparse dictionary model inspired by the SINDy algorithm~\cite{brunton_discovering_2016}.
The latter model builds a library $L$ of mathematical terms, typically low order polynomials of the state and action components, and aims to represent the target observables as a linear combination of these elements. The coefficients for each library element are stored in a weight vector $\mathbf{w}$ such that the predicted observables are given by $o' = L(s, a) \mathbf{w}$. During training, the weights $\mathbf{w}$ are tuned so that the surrogate recovers a correct physical model of the system. The surrogate also helps train the policy $\pi$, which learns to select actions that maximize the expected reward $Q(s, \pi(s))$ and satisfy the \textit{expected} constraints $C[ L(s, \pi(s)) \mathbf{w} ]$. The full algorithms are outlined in Algorithms~\ref{alg:lctd3}-\ref{alg:sindy-lctd3}. 

\begin{figure*}
    \centering
    \includegraphics[width=0.9\linewidth]{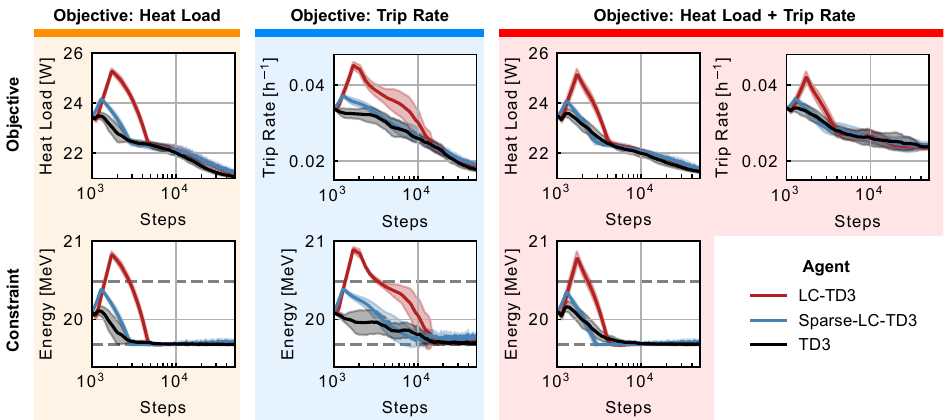}
    \caption{\textbf{Training behavior for RL agents on CEBAF 8D problem}.
    (\textit{Top}) Reward objective at each step for standard TD3 with energy penalty, LC-TD3 (Alg.~\ref{alg:lctd3}) and Sparse LC-TD3 (Alg.~\ref{alg:sindy-lctd3}). 
    Columns denote agents trained to minimize heat load, trip rate, or both simultaneously.
    (\textit{Bottom}) Energy per episode for each agent. Dashed lines denote the bounds of the energy constraint.
    A sparse dictionary surrogate function converges to the energy target faster than a deep neural network.
    Plots show mean and standard deviation over $N=8$ trials per agent per objective.
    }
    \label{fig:training_curve}
\end{figure*}

For each CEBAF optimization environment in Table~\ref{tab:cebaf_problems}, we compared LC-TD3 and Sparse-LC-TD3 (Alg.~\ref{alg:lctd3}-\ref{alg:sindy-lctd3}) to standard TD3 (Alg.~\ref{alg:td3}) for three objectives: minimizing heat load, minimizing trip rate, and minimizing heat load and trip rate simultaneously. The reward functions were 
\begin{align}
    R \in \{ -H, -T, -0.5 (H + T) \}
\end{align}
The energy constraints were chosen following~\cite{rajput_harnessing_2024}
\begin{align}
    C(E) = 
    \begin{cases}
        0 & | E - E_{\text{target}} | < \delta E \\
        -5 \times | E - E_{\text{max}} | & E > E_{\text{target}} + \delta E \\
        -5 \times | E - E_{\text{min}} | & E < E_{\text{target}} - \delta E
    \end{cases}
\end{align}

The discontinuous energy constraint $C(E)$ enables precise control by defining strict boundaries on the energy gain while applying linear penalties when the energy gain is far from the target. 
This helps ensure the agent learns policies that maintain the target energy levels required for optimal accelerator operation.
For the TD3 agents, the energy constraint was integrated into the reward by including a penalty term.
For the LC-TD3 agents, the constraint function was applied to the energy predicted by the learnable surrogate model.

\section{Results}

\subsection{CEBAF test environments}


As an initial step, we trained our agents on a test environment containing 8 cavities, which was a surrogate model for the 1L10 cryo-module. The training performance of each algorithm is shown in Fig.~\ref{fig:training_curve}.
We found that the standard TD3 agent rapidly converged to a solution whose energy gain lay at the lower bound of the allowed energy band. Our Sparse-LC-TD3 agent was able to achieve a similar reward after a longer training cycle, but was quicker to converge than the same algorithm using a neural network surrogate. 
This occured because the neural network model takes time to learn to accurately approximate the energy. 
For both LC-TD3 agents, the reward decreased initially, as the policy was guided by an inaccurate energy surrogate model. 
Once the learned energy surrogate became sufficiently accurate, the agent was able to make informed decisions and reach a high reward. 
When examining the predictions of the energy surrogate, we found that the sparse dictionary model reached the target energy band more quickly than the neural network model (Fig.~\ref{fig:training_curve} bottom).

The Sparse-LC-TD3 agent learned a surrogate model for energy parameterized by coefficients for a set of library functions. 
While the library contained higher-order polynomial terms, we found that the majority of learned coefficients were nearly zero. The identified equation for the energy function contained only linear terms:

\begin{multline}
    E = 20.08 + 0.69\, G_0 + 1.48\, G_1 + 1.55\, G_2 + 0.74\, G_3 \\
    + 0.25\, G_4 + 1.13\, G_5 + 0.90\, G_6 + 1.29\, G_7
    \label{eq:learned_coefs}
\end{multline}

\begin{figure}
    \centering
    \includegraphics[width=0.9\linewidth]{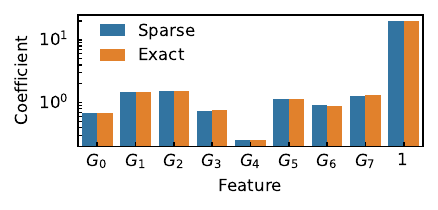}
    \caption{\textbf{Learning a sparse interpretable model of single-cryomodule energy gain.} 
        Comparison of the learned coefficients to their exact values set in the environment.
        Coefficient values were extracted from a Sparse-LC-TD3 agent trained on the CEBAF-8D problem.
        Plot shows coefficients with magnitude greater than a threshold $\tau = 0.05$. 
    }
    \label{fig:coef_comparison}
\end{figure}

This result was expected as the equation for total energy gain (\ref{eq:energy_gain}) contains only terms linear in the cavity gradients.
The identified coefficients accounted for the cavity lengths as well as some features unique to each cavity, such as the maximum and minimum gradient settings allowed by the CEBAF optimization environment. In Fig.~\ref{fig:coef_comparison}, we plotted the identified coefficients next to the ground-truth values from the environment and verified that the machine-learned surrogate learned the correct physics.

\begin{figure}[t]
    \centering
    \includegraphics[width=1.0\linewidth]{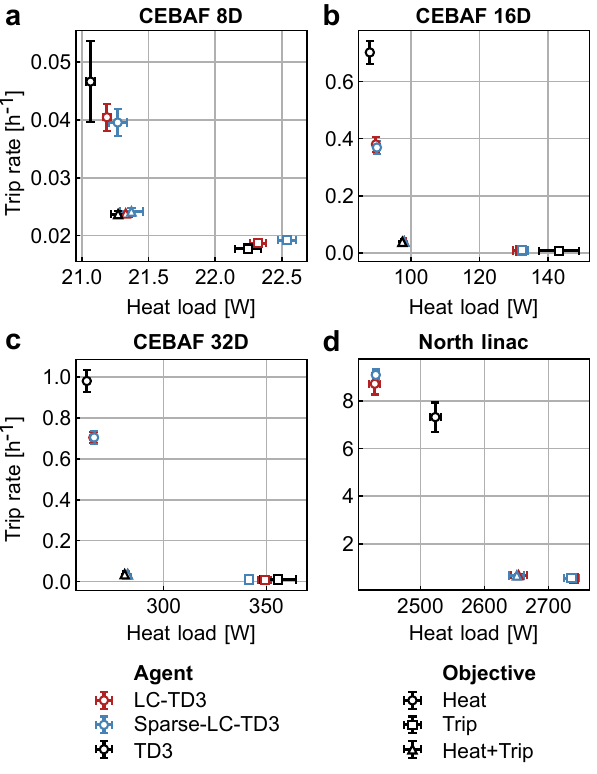}
    \caption{\textbf{Training performance for CEBAF environments}.
        Heat load and trip rate for agents' selected configurations at the end of training for each CEBAF optimization problem. 
        Each training algorithm had $N=8$ total trials for each of the three objectives. 
        We plot the mean and standard deviation for trials which converged to the target energy range. 
        Full results are reported in Table~\ref{tab:linac_north} for North linac and Table~\ref{tab:cebaf8D}-\ref{tab:cebaf32D} for the test environments.
    }
    \label{fig:pareto}
\end{figure}

To assess performance in intermediate-scale problems, we next tested each algorithm on higher-dimensional surrogate environments with 16-dimensional and 32-dimensional state-action spaces. These represented two- and four-cryomodule environments, respectively. 
In Fig.~\ref{fig:pareto}, we plotted the heat load and trip rate selected by the agent for in these environments for each reward function. While TD3 produced more optimal behavior in the one-cryomodule case (Fig.~\ref{fig:pareto}a), this difference disappeared in the two- and four-cryomodule cases (Fig.~\ref{fig:pareto}b-c). All algorithms achieved similar performance on their primary objective in each of these cases.
We did observe that the learnable constraint models achieved more optimal performance on secondary objectives. That is, Sparse-LC-TD3 and LC-TD3 models trained to minimize heat load selected configurations with lower trip rates than standard TD3 models, and (Sparse-) LC-TD3 models trained to minimize trip rate achieved lower heat loads as well. We also found that the learnable constraint models failed to stay within the target energy range for 1-2 trials for each optimization objective, although the majority of trials did converge properly. The complete results are reported in Table~\ref{tab:cebaf8D}-\ref{tab:cebaf32D}.

As in the 8D problem, we examined the learned surrogate by comparing the learned coefficients to their exact values set in the environment. Once again, the majority of identified coefficients were nearly zero and the sparse dictionary model was represented by a linear equation. The non-zero terms exhibited strong agreement with the ground truth values, see Fig.~\ref{fig:intermediate_coef}. This indicated that the sparse surrogate model learned the correct physics governing the energy gain in the accelerator environment.

\subsection{CEBAF North linac environment}

After characterizing our algorithms on small-scale test problems, we turned to the more challenging task of controlling the full CEBAF north linac. This contains 197 SRF cavities whose gradients must be tuned simultaneously. Recent work showed that the TD3 algorithm can struggle to produce configurations near the target energy  for this high-dimensional problem~\cite{rajput_harnessing_2024}. 

We trained RL agents using the TD3 algorithm and both LC-TD3 algorithms for 8 independent trials of 50,000 steps for each objective. 
The performance for each agent and objective are shown in Fig.~\ref{fig:pareto}d and Table~\ref{tab:linac_north}.
We found that our learnable physics-based surrogates enabled agents to select cavity configurations that minimized heat load or FSD trip rate while remaining within the target energy range. 
The LC-TD3 agent proved slightly more consistent than the Sparse-LC-TD3 agent. The former converged to a suitable solution in all trials, while the Sparse-LC-TD3 agent failed to reach the target energy in one trial for two of the objectives. Both LC-TD3 algorithms outperformed standard TD3, which failed to reach the energy target for two objectives and converged only half of the time for its best-performing task. 

\begin{figure}
    \centering
    \includegraphics[width=0.9\linewidth]{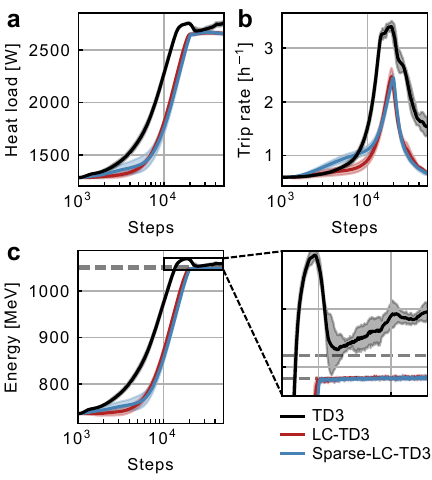}
    \caption{\textbf{Training behavior for RL agents on CEBAF North linac problem}.
        For the predicted configuration at each step, we plot the heat load (\textbf{a}), trip rate (\textbf{b}), and energy gain (\textbf{c}) for each RL agent. Inset zooms in on energy gain near the end of training, showing the failure of TD3 to converge to the target energy range. 
        Plots show mean and standard deviation over $N=8$ agents trained using the combined Heat-Trip objective.
    }
    \label{fig:linac_north_curves}
\end{figure}


In Fig.~\ref{fig:linac_north_curves}, we plotted the heat, trip, and energy for the agents' selected configurations at each step. While the LC-TD3 algorithms reached and remained within the target energy band, the standard TD3 algorithm had trouble adhering to the energy constraints. We hypothesize that this occured due to the high-dimensional nature of the problem. 
Because the volume of the configuration space grows exponentially with the problem dimension, a TD3 agent may require a large number of action samples to characterize the reward landscape near the target energy range. On the other hand, the LC-TD3 and Sparse-LC-TD3 models learned functions that bound the allowed subvolume of configuration space, enabling more efficient sampling and convergence to an optimal solution. 

\begin{table}[t]
\begin{tabular}{l | c || c  c  c}
 & Agent &  Heat (W) &  Trip (h$^{-1}$) & Conv. Rate \\
\hline
\multirow{3}{*}{\rotatebox{90}{Heat}} & LC-TD3 & \textbf{2429 (9)} & 8.72 (0.45) & \textbf{100 \%} \\
 & Sparse-LC-TD3 & \textbf{2429 (5)} & 9.11 (0.21) & 88 \% \\
 & TD3 & 2559 (45) & \textbf{7.08 (0.84)} & 50 \% \\
 \hline
\multirow{3}{*}{\rotatebox{90}{Multi}} & LC-TD3 & 2653 (12) & \textbf{0.67 (0.03)} & \textbf{100 \%} \\
 & Sparse-LC-TD3 & \textbf{2650 (11)} & \textbf{0.67 (0.03)} & 88 \% \\
 & TD3 & 2753 (21) & 1.55 (0.18) & 0 \% \\
 \hline
\multirow{3}{*}{\rotatebox{90}{Trip}} & LC-TD3 & 2740 (6) & \textbf{0.53 (0.03)} & \textbf{100 \%} \\
 & Sparse-LC-TD3 & \textbf{2736 (12)} & 0.55 (0.05) & \textbf{100 \%} \\
 & TD3 & 2784 (20) & 1.09 (0.34) & 0 \% \\
     \end{tabular}
     \caption{ End-of-training performance for each RL agent and objective in the CEBAF North linac environment. Leftmost column indicates the training objective. 
     Mean and standard deviation computed over $N=8$ trials.
     Rightmost column reports the percentage of trials that converged to a configuration producing an energy gain within the allowed range.}
     \label{tab:linac_north}
\end{table}

\begin{figure}
    \centering
    \includegraphics[width=0.7\linewidth]{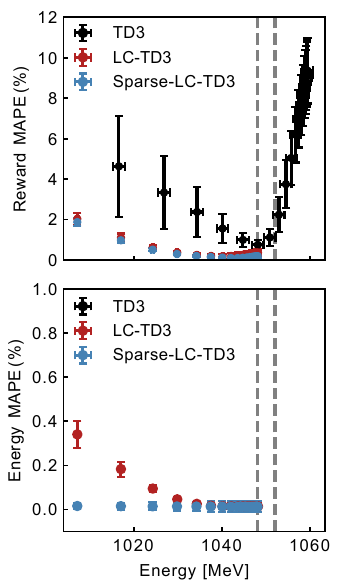}
    \caption{\textbf{Learnable physics surrogate functions enhance performance for high-dimensional action space.}
    (\textit{top}) Relative errors in reward estimation for each RL agent trained on the CEBAF North linac optimization problem.
    (\textit{bottom}) Relative errors in the learned surrogates for LC-TD3 agents. 
    Each point is the average error for randomly-sampled actions a distance $\delta$ from the agents' selected configuration at the end of training, with $\delta$ spaced logarithmically in the range $[0.05, 5]$. 
    Errorbars denote standard deviation over $N=8$ agents trained using the combined Heat-Trip objective.
    }
    \label{fig:error_distance}
\end{figure}

We examined this further by evaluating the local accuracy of the critic and learned surrogate functions. 
To do this, we randomly sampled the action space in the local neighborhood of the trained agents' selected configurations.
By varying the neighborhood size $\delta$ from small to large ($\delta \in [0.05, 5]$), we produced configurations with energy gains near the boundary of the allowed range.
Both LC-TD3 agents more accurately estimated the reward than the TD3 agent. 
The accuracy was most comparable near the target energy, where TD3 approached a 1\% error rate.
However, the LC-TD3 critics remained accurate outside the energy range, while the TD3 critic performed significantly worse (Fig.~\ref{fig:error_distance} top).
Because the TD3 critic was unable to characterize how the energy constraint affected the high-dimensional reward landscape, the policy could not learn to select optimal actions compatible with that constraint. 
By leveraging a learnable energy surrogate model, the LC-TD3 agents could more effectively explore and assess optimal configurations near the bounds of the target energy range. 

We also evaluated the performance of the learned energy surrogates using a similar approach, and found that the sparse dictionary model more accurately estimated the energy gain of new configurations far from the agents' prediction (Fig.~\ref{fig:error_distance} bottom).
This may be beneficial for applications where the experimental or operational objectives may change unexpectedly.
Here, the sparse dictionary model's generalization performance could allow it to assess the physical behavior of a new configuration without requiring costly retraining.

\begin{figure}
    \centering
    \includegraphics[width=0.9\linewidth]{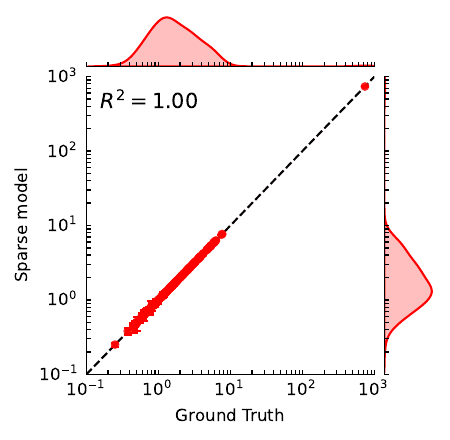}
    \caption{
    \textbf{Learning an accurate model of CEBAF north linac energy gain}.
        Comparison of ground truth and machine-learned coefficient values for the CEBAF north linac environment.
        Values are averaged over all trials of the Sparse-LC-TD3 agent.
    }
    \label{fig:linac_north_coefficients}
\end{figure}

The sparse model agent provides an additional key advantage: interpretability. While the neural network LC-TD3 model could achieve similar performance in minimizing heat load and trip rate and converges slightly more consistently, it gave the operator no insight on how it makes predictions. 
On the other hand, the sparse dictionary model learned a mathematical equation for the energy gain. As in the single-cryomodule case, this equation was a linear combination of terms linear in the cavity gradients $G_i$. 
In Fig.~\ref{fig:linac_north_coefficients}, we compared the inferred values with the ground truths taken from the environment and observed excellent agreement. Thus, an accelerator operator can independently verify that the agent is making decisions based on a correct model of the accelerator physics. 

\subsection{Multi-objective CEBAF optimization}

\begin{figure*}
    \centering
    \includegraphics[width=0.9\linewidth]{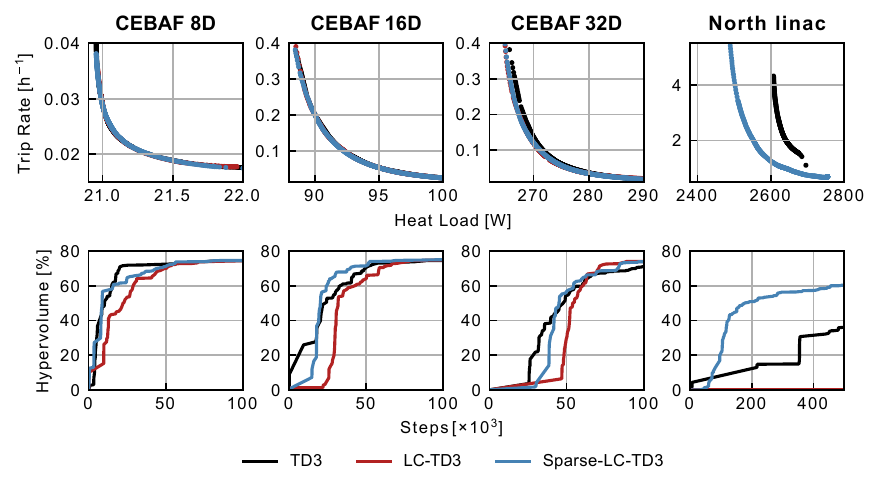}
    \caption{\textbf{Learnable surrogate enables high-dimensional multi-objective optimization.} 
    (\textit{top}) Optimal pareto front produced by each RL agent trained using a multi-objective RL approach.
    (\textit{bottom}) Pareto coverage over time, measured using the normalized hypervolume metric (see Supplement).
    Columns indicate the CEBAF problem being optimized.
    }
    \label{fig:morl_pareto}
\end{figure*}

As a final test of our algorithms' performance, we applied our learnable constraint agents to the conditional multi-objective CEBAF optimization problem considered in~\cite{rajput_harnessing_2024}. In this case, rather than specifying a fixed priority for each objective, we used a tunable weight vector $\alpha$ to set the relative importance of heat load and FSD trip rate minimization. As in~\cite{rajput_harnessing_2024}, we adapted each RL algorithm by using a conditional policy $\pi(s | \alpha)$ and a Q-function $Q_i(s, a)$ that predicts a vector of expected return for each objective. During policy training, the dot product $\alpha \cdot Q$ defines a dynamic scalar optimization objective. This allows the agent to select from Pareto-optimal policies by changing the weight vector $\alpha$. Crucially, the training of the critic and learnable constraint functions are unchanged when considering the multi-objective problem.

\begin{table}[b]
    \begin{tabular}{r || c | c | c | c}
        Agent & 8D & 16D & 32D & North linac \\
        \hline \hline
        TD3 & 74.59 & 74.65 & 71.14 & 36.34 \\
        LC-TD3 & 74.40 & 74.86 & \textbf{74.05} & 0.00 \\
        Sparse-LC-TD3 & \textbf{74.67} & \textbf{75.12} & 73.89 & \textbf{60.42}
    \end{tabular}
    \caption{Pareto front coverage, represented using the normalized hypervolume metric. Reference and ideal points for hypervolume calculation are listed in Table~\ref{tab:hv_points}.}
    \label{tab:morl_pareto}
\end{table}

We trained RL agents using each algorithm and evaluated the quality of predicted configurations at 500 evenly-spaced weight vectors. We trained agents for 100,000 steps for the 8D, 16D, and 32D problem and 500,000 steps for the North linac to give agents more time to converge on the high-dimensional problem. To quantify relative performance, we used the normalized hypervolume metric~\cite{guerreiro_hypervolume_2021} as done in~\cite{rajput_harnessing_2024}, see Supplement for details. Briefly, the normalized hypervolume describes the size of the objective space captured by the agent's Pareto front. We computed hypervolume with respect to predicted configurations within the target energy range for each problem.

The conditional policy predictions and normalized hypervolumes for each problem are plotted in Fig.~\ref{fig:morl_pareto} and reported in Table~\ref{tab:morl_pareto}. 
For the 8D and 16D problems, the algorithms were nearly indistinguishable. For the 32D problem, the LC-TD3 and Sparse LC-TD3 algorithms achieved 3.9\% and 4.1\% improvements over standard TD3. In the North linac case, the standard TD3 model struggled to generated predictions within the target energy range.
Less than half of its predicted configurations satisfied the energy constraint, resulting in a less-populated Pareto front (see Fig.~\ref{fig:morl_pareto} top). 
LC-TD3 also failed to satisfy the energy constraint. This likely occurred because the surrogate model, represented by a neural network, did not accurately characterize the broad high-dimensional energy landscape needed to generate valid configurations at different objective priorities. Even in the single-objective case, we found that LC-TD3 models struggled to capture the relevant physics outside a narrow region near the selected configuration (Fig.~\ref{fig:error_distance}). 

Sparse LC-TD3 achieved the highest performance for the North linac problem, generating a continuum of configurations within the target energy range at different prioritization levels (Fig.~\ref{fig:morl_pareto}). The learned constraint represented by a sparse dictionary model better characterized the energy over the high-dimensional state-action space, enabling the agent to make informed decisions. 
This sparse dictionary model did not require a differentiable environment. Instead, it learned a parsimonious representation of the relevant physics from observations that in turn guided its policy optimization. 

\begin{figure}[t]
    \centering
    \includegraphics[width=0.9\linewidth]{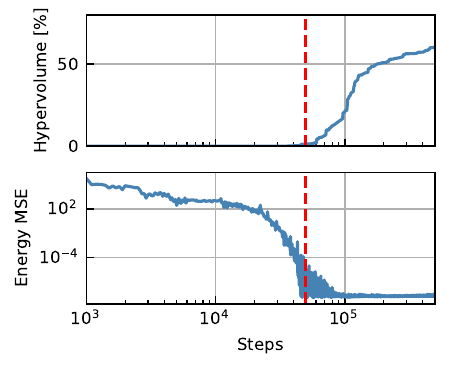}
    \caption{\textbf{Sparse-LC-TD3 surrogate accuracy and predictive performance for North linac problem.}
        (\textit{Top}) Pareto coverage at each episode step.
        (\textit{Bottom}) Error rate of energy surrogate at each episode step.
        Red line denotes step 50,000 after which the energy surrogate began to stabilize and the predictions improved.
    }
    \label{fig:surrogate_accuracy}
\end{figure}

We note that for the North linac, the agent's energy model required significant time to stabilize and this additional warm-up period slowed the overall optimization. 
The predicted cavity configurations did not begin improving until after the energy model converged and the only started to plateau after 250,000 training steps (Fig.~\ref{fig:surrogate_accuracy}). 
The RL agent studied in~\cite{rajput_harnessing_2024} benefited from an accurate and fully differentiable surrogate environment from the start of training. 
As a result, it achieved faster convergence and generated predictions with lower heat loads and higher Pareto coverage. 
A focus of future work will be close this gap by speeding up the surrogate training process, potentially via library pruning, coefficient regularization, or alternative optimizers. 

\section{Conclusion}

We presented a RL framework that learns physical constraints from environmental observables in order to determine optimal control configurations for particle accelerator environments. We demonstrated this approach on surrogate models of the CEBAF facility at Jefferson Lab and found that it outperformed a traditional RL algorithm on a high-dimensional experimental optimization task.
Our procedure blends model-free and model-based RL techniques. 
The agent does learn a model of relevant physical observables, such as energy, which plays a crucial role in shaping the policy. 
However, the agent does not attempt to predict the complete dynamics of the full environment, which can be much more complicated and are less likely to be represented by a simple mathematical rule (see Fig.~\ref{fig:pendulum-training}-\ref{fig:pendulums-coefs} for an example using the classic pendulum control problem). 
These results corroborate and extend recent work showing how using a differentiable physics-based surrogate environment improves RL performance on accelerator tasks by allowing gradient back-propagation~\cite{rajput_harnessing_2024}.
Here, we showed how this surrogate can be learned from the data during agent training, bypassing the time-consuming task of building differentiable environments for complex particle accelerators.
The surrogate also makes the agent more interpretable. An operator can probe the learned equation and verify it against known physics of the system.
This grey-box machine learning approach is a promising path towards discovery~\cite{colen_interpreting_2024,lefebvre_learning_2024,champion_data-driven_2019}, characterization~\cite{seara_sociohydrodynamics_2024,schmitt_machine_2024,murphy_machine-learning_2024,murphy_information_2024,hanlan_cornerstones_2024}, and control~\cite{zolman_sindy-rl_2024,falk_learning_2021,floyd_tailoring_2024} of complex experimental systems whose governing physics may be obscured or unknown.


\section{Acknowledgments}
\begin{acknowledgments}
Funding for this effort was provided by The Hampton Roads Biomedical Research Consortium as part of the efforts associated with the Joint Institute for Advanced Computing on Environmental Studies between Old Dominion University and Jefferson Laboratory. 
This research was supported by the Research Computing Clusters at Old Dominion University. 
This manuscript has been co-authored by Jefferson Science Associates (JSA), operating the Thomas Jefferson National Accelerator Facility for the U.S. Department of Energy under Contract No. DE-AC05-06OR23177.
\end{acknowledgments}

\bibliographystyle{apsrev4-1}
\bibliography{references.bib}

\clearpage
\newpage
\newpage

\setcounter{figure}{0}
\renewcommand{\thefigure}{S\arabic{figure}}

\setcounter{table}{0}
\renewcommand{\thetable}{S\arabic{table}}

\setcounter{algocf}{0}
\renewcommand{\thealgocf}{S\arabic{algocf}}

\setcounter{equation}{0}
\renewcommand{\theequation}{S\arabic{equation}}

\section{Supplementary Information}

\begin{algorithm}
\caption{Twin-delayed deep deterministic policy gradient (TD3)}
\label{alg:td3}
\SetAlgoLined
Initialize critics $Q_{\theta_1}, Q_{\theta_2}$ and policy network $\pi_{\phi}$ \\
Initialize target networks $\theta_{i'} \gets \theta_{i}$, $\phi' \gets \phi$ \\
Initialize replay buffer $\mathcal{B}$ \\
\For{$e$ in $1\dots N_e$}{
    Observe state $s$ and select action $a \sim \pi_{\phi}$ \\
    Execute $a$ in environment \\
    Observe next state $s'$, reward $r$, and terminal signal $d$ \\
    Store $(s, a, r, s', d)$ in replay buffer $\mathcal{B}$ \\
    \If{time to update}{
        Sample batch of transitions $b \sim \mathcal{B}$ \\
        $a' \gets \pi_{\phi'}(s') + \epsilon\, \mathcal{N}(0, \sigma)$ \\
        $y_i \gets r + \gamma \min_i Q_{\theta'_i}(s', a')$ \\
        Update $Q$ functions with gradient descent using 
        $\frac{1}{|b|} \nabla_{\theta_i} \sum \left( Q_{\theta_i}(s, a) - y_i \right)^2$ \\
        Update policy $\pi$ with gradient ascent using 
        $\frac{1}{|b|} \nabla_{\phi} \sum Q_{\phi_1}(s, \pi_{\phi}(s))$ \\
        Update target networks: \\
        $\theta_i' \gets \tau \theta_i' + (1 - \tau) \theta_i $ \\
        $ \phi' \gets \tau \phi' + (1 - \tau) \phi$
    }
}
\end{algorithm}

\subsection{Quantifying Pareto front coverage for multi-objective RL}

\begin{figure}
    \centering
    \includegraphics[width=0.8\linewidth]{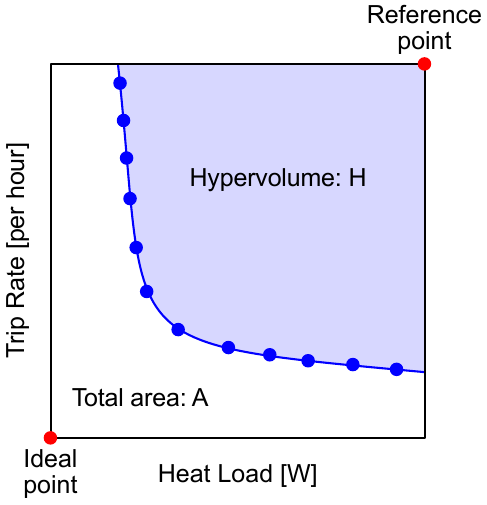}
    \caption{Hypervolume calculation for a 2D Pareto front using chosen reference an ideal points.}
    \label{fig:hypervolume}
\end{figure}

To quantify the performance of RL agents in the multi-objective problem, we used the normalized hypervolume or S-metric~\cite{guerreiro_hypervolume_2021} approach as in ~\cite{rajput_harnessing_2024}. For this metric, an ideal and reference point are chosen for each objective value. These represent an upper and lower bound respectively for the region of objective space. The hypervolume $H$ is the volume of objective space delineated by the Pareto front. The normalized hypervolume reported in Table~\ref{tab:morl_pareto} is given by
\begin{equation}
    NH = H / A \times 100 
\end{equation}
Here $A$ is the total volume of the objective region defined by the reference and ideal points. Figure~\ref{fig:hypervolume} visualizes this calculation for the multi-objective problem of minimizing heat load and trip rate. The normalized hypervolumes reported in this paper were computed using the Pymoo package~\cite{blank_pymoo_2020}, with respect to the ideal and reference points given in Table~\ref{tab:hv_points}. 
For the CEBAF 8D, 16D, and 32D problems, the reference and ideal points are identical to those used in~\cite{rajput_harnessing_2024}. 

For the North linac, we selected new upper and lower bounds to enable a clearer comparison between Sparse-LC-TD3 and standard TD3, as the latter produced configurations above the original reference point leading to a normalized hypervolume of 0. For the DDRL agent results reported in~\cite{rajput_harnessing_2024}, we calculated a normalized hypervolume of 77.79 with respect to the bounds in Table~\ref{tab:hv_points}.

\begin{table}[]
    \centering
    \begin{tabular}{c| c | c}
         Problem &  Ref (H, T) & Ideal (H, T) \\
         \hline 
         \hline
         8D & (22.4, 0.05) & (20.9, 0.015) \\
         16D & (100, 0.40) & (88, 0.015) \\
         32D & (290, 0.40) & (262, 0.01) \\
         North linac & (2800, 5.5) & (2380, 0.5) 
    \end{tabular}
    \caption{Reference and ideal points used for multi-objective hypervolume calculation.}
    \label{tab:hv_points}
\end{table}

\subsection{Benchmark on OpenAI-Pendulum}

\begin{figure}[b]
    \centering
    \includegraphics[width=0.8\linewidth]{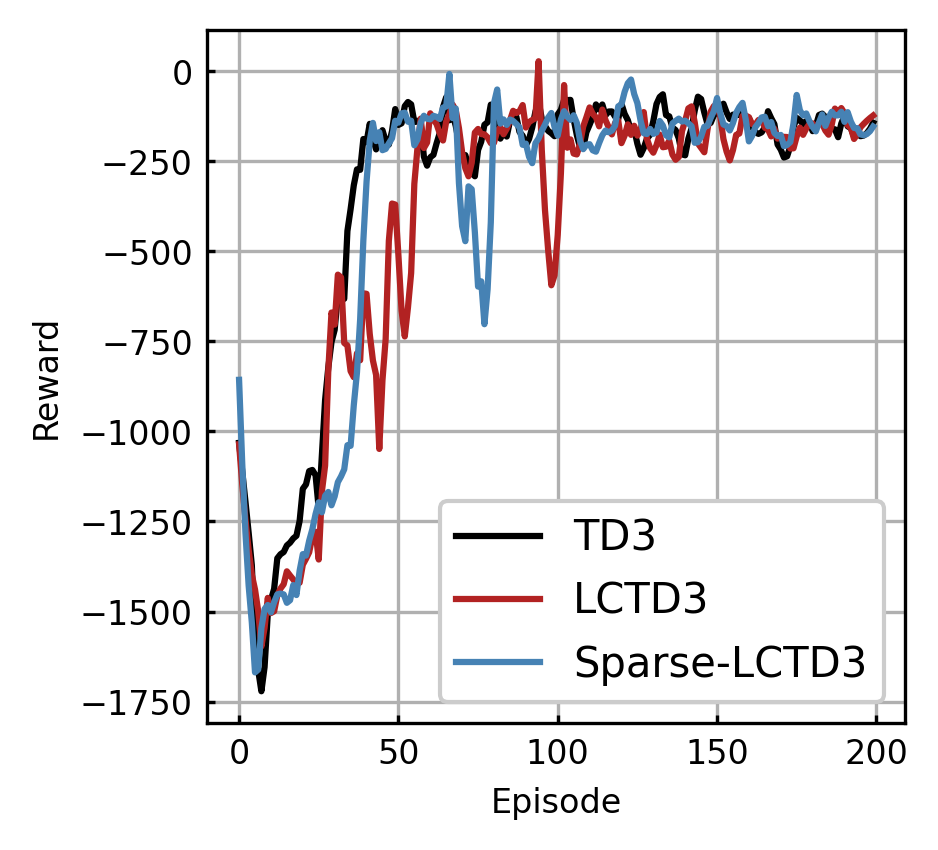}
    \caption{Training curves for TD3, LC-TD3, and Sparse-LC-TD3 agents trained on a Pendulum-v1 environment}
    \label{fig:pendulum-training}
\end{figure}

To demonstrate our algorithm's performance on a more standard reinforcement learning environment, we consider here the Pendulum-v1 environment from OpenAI-Gym~\cite{brockman_openai_2016}. The system contains a free pendulum subject to gravity. The agent observes a state $s = (x, y, \omega)$ where $(x, y)$ is the position of the pendulum end and $\omega$ is the angular velocity. The objective is for the agent to continuously apply a torque $\tau$ such that the pendulum swings upright and remains inverted. We add the additional constraint that the kinetic energy remain below some specific value $\frac{1}{2} I \omega^2 \le T_{\text{max}}$.

We trained RL agents using Algorithms~\ref{alg:lctd3}-\ref{alg:sindy-lctd3} for 200 episodes with $T_{\text{max}} = 2$ and plotted the reward after each episode in Fig.~\ref{fig:pendulum-training}. As a comparison, we also plot the reward for a TD3 agent (Alg.~\ref{alg:td3}) with no energy penalty ($T_{\text{max}} \rightarrow \infty$). After training, the two LC-TD3 agents achieve comparable rewards to the standard TD3 agent.

\begin{figure}
    \centering
    \includegraphics[width=0.9\linewidth]{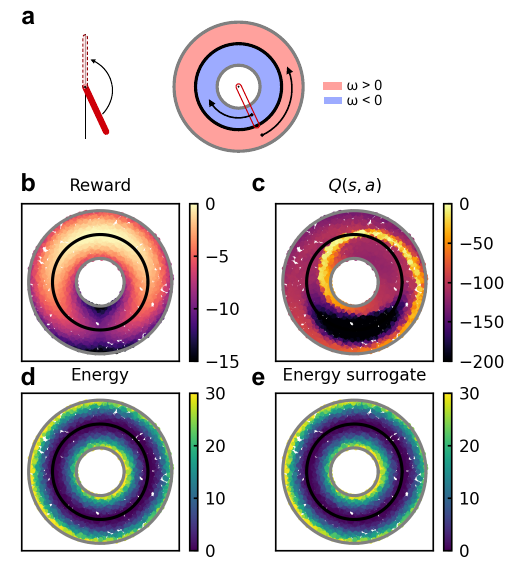}
    \caption{
        (\textbf{a}) The Pendulum-v1 objective is to swing the pendulum upright. We plot the state space such that the angular coordinate shows the pendulum position and the radial coordinate denotes the angular velocity.
        (\textbf{b}) Ground truth reward landscape for the pendulum environment.
        (\textbf{c}) Learned critic $Q(s, a)$ is a nonlinear function of state and action variables.
        (\textbf{d}) Ground truth energy function.
        (\textbf{e}) Energy surrogate model.
        All plots show ground truth and predictions for $N=5000$ randomly sampled state-action pairs. 
    }
    \label{fig:pendulum-space}
\end{figure}

In Fig.~\ref{fig:pendulum-space}, we examined the agent's learned behavior and compared it to the known physical behavior of the pendulum environment.
The reward function was given exactly by
\begin{align}
    R = -\theta^2 - \alpha \omega^2 - \beta \tau^2
\end{align}
The critic's learned Q-function was more complicated and depended nonlinearly on the pendulum position, the direction of motion, and the direction of the applied torque. It was difficult to represent this function using a simple equation. The energy function was comparatively simple and the learned surrogate is accurate to the exact values given by the environment.

\begin{figure}[ht]
    \centering
    \includegraphics[width=0.9\linewidth]{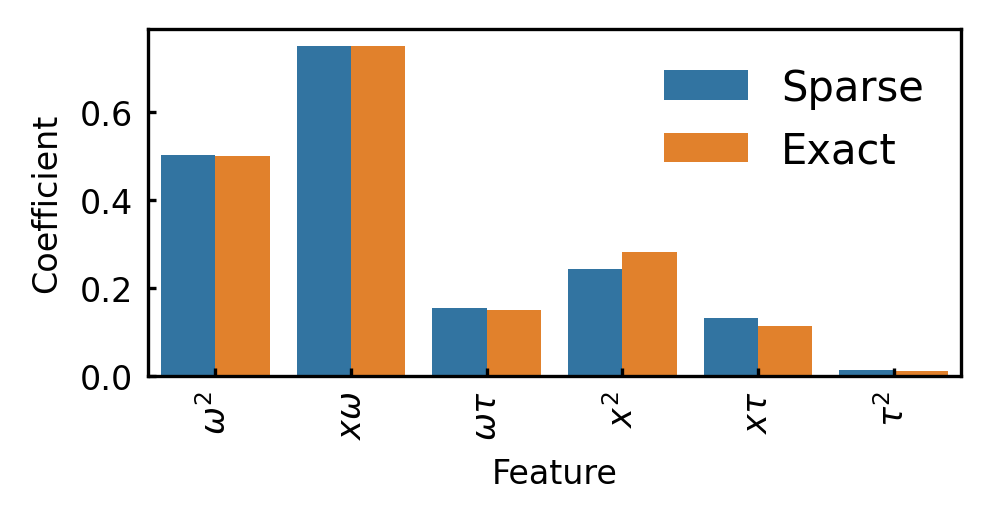}
    \caption{Sparse-LC-TD3 energy surrogate coefficients for Pendulum-v1 environment compared to the exact result of Eq.~\ref{eq:pendulum_coefs}}
    \label{fig:pendulums-coefs}
\end{figure}

For the Sparse-LC-TD3 agent, we compared the learned surrogate model to the exact equation for energy within the environment. Recall that $o$ is returned by the environment after performing an action, so the model predicts the new kinetic energy after the angular velocity updates. The ground truth equation is
\begin{align}
    E(t + \Delta t) = \frac{1}{2} I \left[ \omega + \left( \frac{3 g}{2 \ell} x + \frac{3}{m \ell^2} \tau \right) \Delta t \right]^2
    \label{eq:pendulum_coefs}
\end{align}
Here $I$ is the moment of inertia, $g$ is the acceleration due to gravity, $m$ is the pendulum mass, $\ell$ is the pendulum length, and $\Delta t$ is the simulation time step. The term in brackets is the angular velocity after one time step, which can be obtained by inspecting the open-source environment code or by manual derivation. In Fig.~\ref{fig:pendulums-coefs}, we compared the numerical values for each coefficient in this equation to the corresponding learned values in the weight vector $\mathbf{w}$. The agent learned coefficients that closely align with the ground truth. Examining the constraint surrogate in this way allows one to ensure that the policy is shaped by an accurate representation of the relevant physics of the system. 

\clearpage
\onecolumngrid

\subsection{CEBAF test environment results}

\begin{table*}[h]
\begin{tabular}{l  l || c  c  c}
Objective & Agent &  Heat load (W) &  Trip rate (h$^{-1}$) & Convergence \\
\hline
\hline
Heat & LC-TD3 & 21.18 (0.04) & 0.0410 (0.0026) & 88 \% \\
 & Sparse-LC-TD3 & 21.27 (0.07) & 0.0395 (0.0023) & 100 \% \\
 & TD3 & 21.06 (0.03) & 0.0466 (0.0070) & 100 \% \\
\hline
Multi & LC-TD3 & 21.32 (0.04) & 0.0236 (0.0006) & 75 \% \\
 & Sparse-LC-TD3 & 21.37 (0.09) & 0.0242 (0.0007) & 100 \% \\
 & TD3 & 21.27 (0.05) & 0.0237 (0.0006) & 100 \% \\
\hline
Trip & LC-TD3 & 22.32 (0.06) & 0.0187 (0.0004) & 100 \% \\
 & Sparse-LC-TD3 & 22.54 (0.07) & 0.0192 (0.0004) & 100 \% \\
 & TD3 & 22.24 (0.10) & 0.0177 (0.0004) & 100 \% \\
     \end{tabular}
     \caption{ End-of-training performance for each RL agent and objective in the single-cryomodule 8D optimization problem. Mean and standard deviation computed over $N=8$ trials. Convergence denotes the percentage of trials that converged to a configuration producing an energy gain within the allowed range.}
     \label{tab:cebaf8D}
\end{table*}

\begin{table*}[h]
\begin{tabular}{l  l || c  c  c}
Objective & Agent &  Heat load (W) &  Trip rate (h$^{-1}$) & Convergence \\
\hline
\hline
Heat & LC-TD3 & 89.7 (0.1) & 0.3796 (0.0261) & 100 \% \\
 & Sparse-LC-TD3 & 90.0 (0.3) & 0.3651 (0.0212) & 75 \% \\
 & TD3 & 88.0 (0.2) & 0.7013 (0.0402) & 100 \% \\
\hline
Multi & LC-TD3 & 98.0 (0.4) & 0.0398 (0.0014) & 88 \% \\
 & Sparse-LC-TD3 & 98.0 (0.2) & 0.0393 (0.0008) & 100 \% \\
 & TD3 & 97.5 (0.4) & 0.0386 (0.0013) & 100 \% \\
\hline
Trip & LC-TD3 & 131.8 (1.9) & 0.0081 (0.0002) & 100 \% \\
 & Sparse-LC-TD3 & 132.5 (1.6) & 0.0088 (0.0002) & 100 \% \\
 & TD3 & 143.3 (5.8) & 0.0072 (0.0004) & 100 \% \\
     \end{tabular}
     \caption{ End-of-training performance for each RL agent and objective in the two-cryomodule 16D optimization problem. Mean and standard deviation computed over $N=8$ trials. Convergence denotes the percentage of trials that converged to a configuration producing an energy gain within the allowed range.}
     \label{tab:cebaf16D}
\end{table*}

\begin{table*}[h]
\begin{tabular}{l  l || c  c  c}
Objective & Agent &  Heat load (W) &  Trip rate (h$^{-1}$) & Convergence \\
\hline
\hline
Heat & LC-TD3 & 265.7 (0.5) & 0.7017 (0.0249) & 88 \% \\
 & Sparse-LC-TD3 & 266.0 (0.8) & 0.7086 (0.0301) & 88 \% \\
 & TD3 & 262.6 (0.2) & 0.9804 (0.0545) & 100 \% \\
\hline
Multi & LC-TD3 & 282.5 (0.6) & 0.0344 (0.0012) & 100 \% \\
 & Sparse-LC-TD3 & 282.3 (0.6) & 0.0353 (0.0014) & 88 \% \\
 & TD3 & 281.0 (0.7) & 0.0352 (0.0012) & 100 \% \\
\hline
Trip & LC-TD3 & 349.2 (3.2) & 0.0083 (0.0005) & 100 \% \\
 & Sparse-LC-TD3 & 341.7 (2.0) & 0.0100 (0.0003) & 100 \% \\
 & TD3 & 355.8 (8.9) & 0.0096 (0.0015) & 100 \% \\
     \end{tabular}
     \caption{ End-of-training performance for each RL agent and objective in the four-cryomodule 32D optimization problem. Mean and standard deviation computed over $N=8$ trials. Convergence denotes the percentage of trials that converged to a configuration producing an energy gain within the allowed range.}
     \label{tab:cebaf32D}
\end{table*}

\begin{figure*}
    \centering
    \includegraphics[width=0.9\linewidth]{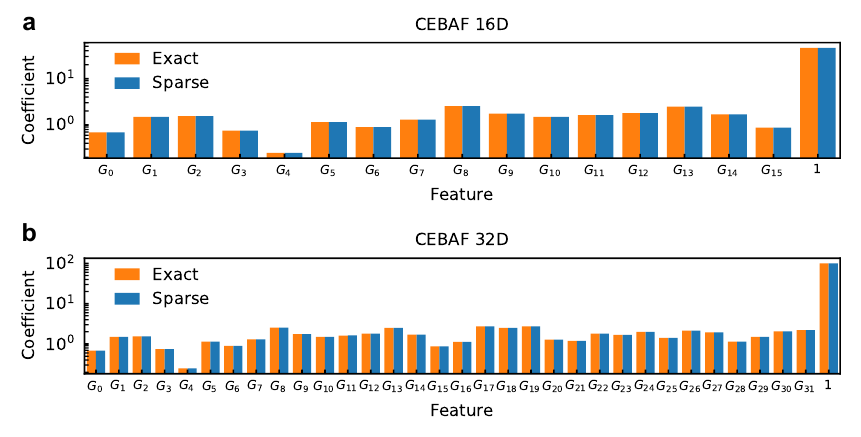}
    \caption{ Comparison of average learned coefficients compared to their values set in intermediate-scale CEBAF environments. 
        (\textbf{a}) Results for CEBAF 16D environment.
        (\textbf{b}) Results for CEBAF 32D environment.
        Plots show all coefficients with magnitudes above a threshold $\tau = 0.05$. Coefficient values are averaged over all trials of the Sparse LC-TD3 agent.
    }
    \label{fig:intermediate_coef}
\end{figure*}

\end{document}